\documentclass[amsmath,amssymb,aps]{revtex4-2}

\usepackage{graphicx}
\usepackage{dcolumn}
\usepackage{bm}
\usepackage{amsmath}

\usepackage{diagbox}
\usepackage{pifont}
\usepackage{multirow}
\usepackage{ctable}
\usepackage{float}

\usepackage{epstopdf}
\begin{document}

\preprint{APS/123-QED}









\title{Compression Spectrum: Where Shannon meets Fourier}


\author{Aditi Kathpalia}
\email{kathpalia@cs.cas.cz}
 \affiliation{%
 Department of Complex Systems, \\
 Institute of Computer Science of the Czech Academy of Sciences, \\
 Prague, Czech Republic
}
 \author{Nithin Nagaraj}%
\email{nithin@nias.res.in}
\affiliation{%
 Consciousness Studies Programme, National Institute of Advanced Studies, 
 Indian Institute of Science Campus,\\ 
 Bengaluru 560 012, India
}%






\maketitle


\section*{Abstract}
Signal processing and Information theory are two disparate fields used for characterizing signals for various scientific and engineering applications. Spectral/Fourier analysis, a technique employed in signal processing, helps estimation of power at different frequency components present in the signal.  Characterizing a time-series based on its average amount of information (Shannon entropy) is useful for estimating its complexity and compressibility (eg., for communication applications). Information theory doesn't deal with spectral content while signal processing doesn't directly consider the information content or compressibility of the signal. In this work, we attempt to bring the fields of signal processing and information theory together by using a lossless data compression algorithm to estimate the amount of information or `compressibility' of time series at different scales. To this end, we employ the Effort-to-Compress (ETC) algorithm to obtain what we call as a \emph{Compression Spectrum}. This new tool for signal analysis is demonstrated on synthetically generated periodic signals, a sinusoid, chaotic signals (weak and strong chaos) and uniform random noise. The Compression Spectrum is applied on heart interbeat intervals (RR) obtained from real-world normal young and elderly subjects. The compression spectrum of healthy young RR tachograms in the log-log scale shows behaviour similar to $1/f$ noise whereas the healthy old RR tachograms show a different behaviour. We envisage exciting possibilities and future applications of the Compression Spectrum.

\noindent {\bf Keywords --} Spectral analysis, Information content, Compression Spectrum, Effort-To-Compress, Compression-Complexity, $1/f$  noise

\section{\label{sec:level1}Introduction}

Signal processing~\cite{oppenheim1975digital} and information theory~\cite{shannon1948mathematical, cover1999elements} are two disparate fields used for characterizing signals for various scientific and engineering applications. Spectral analysis or frequency domain analysis is an important area of signal processing that helps to estimate the power at different frequency components present in the signal, e.g., for signal processing applications. In case of information theory, characterizing data (time-series, text, audio, video signals etc.) based on its average amount of information is useful for estimating its complexity and compressibility, e.g., for communication applications. Information theory considers observations as random variables and does not directly care about the signal aspect (such as power, frequency components) of time series. On the other hand, signal processing does not explicitly take into account the complexity, information content and compressibility of the signal.

Data compression~\cite{sayood2017introduction, cover1999elements} is closely tied to information theory and allows for an algorithmic approach to estimate complexity and compressibility of time series. Information entropy of a random variable provides limits on the amount of its compressibility~\cite{shannon1948mathematical, cover1999elements}. In this work, we attempt to bring the fields of signal processing and information theory together by using a lossless data compression algorithm to estimate the amount of information or `compressibility' of time series at different scales or frequencies. We use the Effort-to-Compress (ETC) algorithm~\cite{nagaraj2013new} to develop this idea of the \emph{Compression Spectrum}. The compression spectrum is obtained for periodic, chaotic and random simulated processes. Subsequently, we also demonstrate its application to real world data, particularly heartbeat RR tachograms obtained from healthy young and old subjects.

\section{Methods}

The compression spectrum is obtained using the Effort-To-Compress (ETC) algorithm~\cite{nagaraj2013new}. ETC is a measure of \emph{compression-complexity}~\cite{nagaraj2017three}. Measures of compression-complexity are used to estimate the complexity of a time series based on lossless data compression algorithms. ETC employs the Non Sequential Recursive Pair Substitution algorithm~\cite{ebeling1980grammars}. Given time series is first converted to a symbolic sequence~\footnote{A symbolic sequence is a sequence consisting of symbols for each value of the time series, obtained by quantization. For example, a binary symbolic sequence would be one where every value of the time-series is replaced by `$0$' or `$1$' based on some threshold. One can have any finite number of bins for quantizing the input time series. Bins could be uniform or not. Typically, uniform bins are chosen to create the symbolic sequence from the time-series.} and then parsed from left to right to find the most frequently occurring pair in the sequence. This pair is then replaced by a new symbol (one that is not in the alphabet of the symbolic sequence). This procedure of finding the most frequently occurring pair is repeated iteratively until the length of the sequence becomes $1$ or all the symbols in the sequence become the same. At this point the Shannon entropy of the sequence becomes $0$ bits. To give an example, ETC algorithm when applied on the input symbolic sequence `1121122112' will transform it as follows: $1121122112 \rightarrow 3232232 \rightarrow 4424 \rightarrow 445 \rightarrow 65 \rightarrow 7$. 

In order to understand how the compression spectrum is obtained, we introduce the idea of \emph{scales} here. Scale of a selected pair of symbols (a pattern) at any iteration of ETC is basically the length of the pattern (in the original sequence). To make this clear, let us see which patterns correspond to which scales in the example taken above. Here, `1’ or `2’ is a pattern corresponding to scale 1. Since ETC finds patterns of length 2 or higher, there is no compression taking place at scale 1. `11’, `12’, `21' or `22' are patterns corresponding to scale 2. After the first ETC iteration, higher order patterns can be seen. `32’ is a pattern of scale 3 as it combines a pattern of scale 2 (`3’ itself now corresponds to scale 2 as it is a representation of `11’) and pattern of scale 1. Further down, `4’ is a symbol of scale 3 and hence the pair `24’ belongs to the $4^{th}$ scale. This way a scale can be attributed to each pair selected in ETC iterations. 

A couple of things to note in the working of the ETC algorithm when compression spectrum is estimated are as follows. When more than one pair has maximum frequency of occurrence in the sequence, then the pair corresponding to the shortest scale is chosen and substituted. In case of the example taken above, this is the reason that the pair `24' is chosen for substitution (instead of the pair `44') in the third iteration of the algorithm. Furthermore, only those iterations of the ETC algorithm are considered in the estimation of the spectrum until which all the pairs in the sequence become unique and subsequent iterations are ignored. In the above example, the algorithm's steps only up to the second iteration will be used in the estimation of the compression spectrum (since the sequence `4424' has pairs `44', `42' and `24' - which are all unique and occur only once each). The rationale for this criteria is that once a sequence has been iteratively  transformed to contain pairs having unique pairs (with a single occurrence each), it has essentially been reduced to a completely {\it random} sequence having the maximum compression-complexity (as estimated by the algorithm going forward). The patterns that characterized the original sequence have all been {\it squeezed} (compressed or learnt) into symbols creating higher scale patterns, and nothing remains to be squeezed or compressed. What remains are symbols (typically that correspond to higher scales) that are making up a completely random sequence. Subsequent iterations of the algorithm are not learning anything important since all pairs of this random sequence are unique and have a frequency of occurrence equal to 1. 

At each iteration, notice that the length of the sequence always reduces. Thus, a compression ratio can be clearly defined and computed at each iteration. This compression ratio ($CR_i$) for any iteration $i$ is defined as follows:
\begin{equation}
    CR_i=\frac{\text{length before pair substitution}}{\text{length after pair substitution}}.
\end{equation}

Further, the compression ratio corresponding to a particular scale $s$ can be given by the fraction as defined above for all iterations corresponding to this particular scale put together. Since these ratios computed at different ETC iterations are multiplicative for a particular $s$, they will be additive in the logarithmic scale. Thus,

\begin{equation}
    \log_2{CR(s)}=\sum_i{\log_2{CR_i}},
\end{equation}

for all iterations $i$ where the pair substituted corresponds to scale $s$. {\it Compression Spectrum} is defined as the plot of the compression ratios ($CR$) at different scales vs. scale.

\section{Results}
\label{section_res}
\subsection{Simulations}

\begin{figure}
\begin{center}
\includegraphics[width=0.8\linewidth]{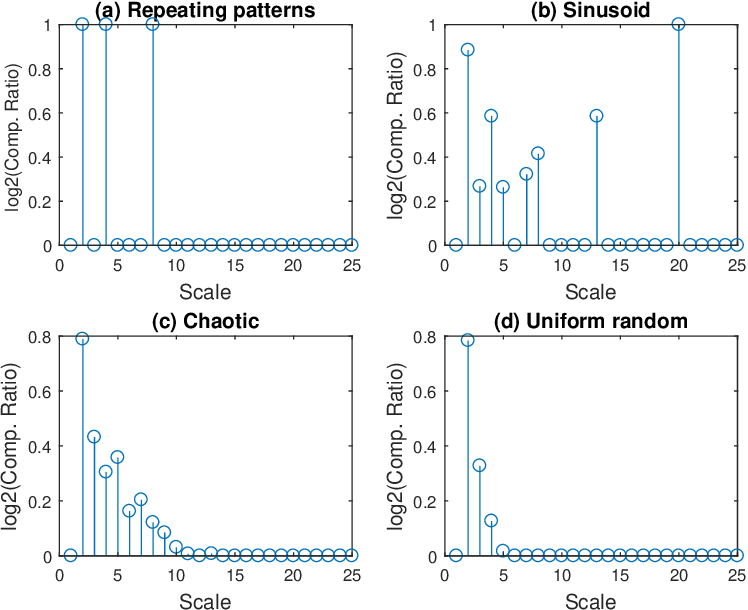}
  \caption{Compression spectra for different types of signals.}
\label{fig1}
\end{center}
\end{figure}

The compression spectrum for a number of simulated signals (length$=2000$ and $8$ bins for quantizing the signal to obtain the symbolic sequence) is shown in Fig.~\ref{fig1}.  Fig.~\ref{fig1}(a) displays the spectrum for a signal $X$ composed of repeating patterns of length 8: $[1,2,3,4,5,6,7,8]$. Similarly, Fig.~\ref{fig1}(b) shows the spectrum for $X(t)=\sin{(2\pi w t)}$, where $w=50$ Hz and sampling frequency is $1$ kHz; \ref{fig1}(c) is the spectrum for the time series from the chaotic logistic map: $X(n)=aX(n-1)(1-X(n-1))$, where $a=4.0$; and \ref{fig1}(d) is for a noise signal comprised of entries drawn from a uniform random distribution $U(0,1)$. 

In Fig.~\ref{fig1}(a), compressibility at patterns corresponding to scales 2, 4 and 8 can be clearly observed. In \ref{fig1}(b), the highest scale (at which compressibility is also maximum) can be observed at 20, which is the fundamental period (number of samples in one cycle) of the sinusoid. There are other lower scales at which compressibility is non-zero as there is some periodicity at those scales. Further, both \ref{fig1}(c) and \ref{fig1}(d) have a decaying spectrum as there is no periodicity in the signal and much of the information content is concentrated in lower scales. However, the spectrum of (c) is broader than that of (d).  This is because, a random noise signal is expected to have negligible information or compressibility at any specific scale. The spectrum obtained at lower scales is due to the repeatability of patterns by chance.  

%
%
%
%
%
\begin{figure}[h]
\begin{center}
\includegraphics[width=0.8\linewidth]{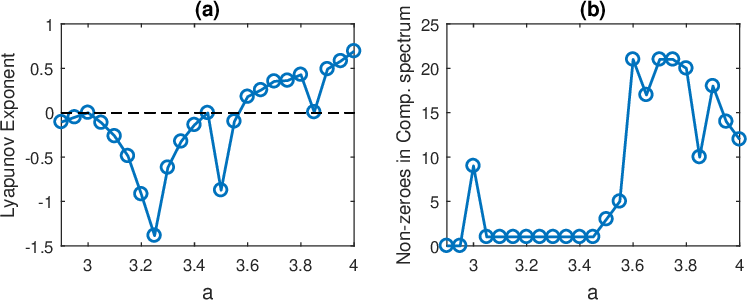}
  \caption{(a) Lyapunov exponent and (b) number of non-zero values in the compression spectrum of logistic maps for varying parameter $a$.}
\label{fig2}
\end{center}
\end{figure}

Further, we analyse the variation of compression spectra as the level of chaos is varied in case of a logistic map time series. This was simulated in the same manner as for Fig.~\ref{fig1}(c) above and the bifurcation parameter $a$ was varied from $2.9$ to $4.0$. Fig.~\ref{fig2}(a) displays Lyapunov exponent with varying $a$ while \ref{fig2}(b) shows the number of non zero values in the corresponding compression spectrum. Here, we can notice that the compression spectrum is broadest at intermediate levels of chaos. Akin to fourier analysis, we could define the concept of {\it bandwidth} as the number of scales at which the compression ratio is non-zero. As it can be seen from~\ref{fig2}(b), the bandwidth is maximum for {\it weak chaos}. 

\subsection{Real data}

We also analyse time series of heart interbeat intervals recorded from young and elderly healthy human subjects using the compression spectrum. These were take from the `Physionet: Fantasia database'~\cite{goldberger2000physiobank}. The original recordings were made for~\cite{iyengar1996age}. In their study, continuous ECG data were recorded from the subjects while watching the movie Fantasia (Disney, 1940) in supine resting state. These recordings were then digitized by sampling at 250 Hz. The occurrence of each ``R" peak was noted, and the time series consisting of the time difference between successive peaks was generated. This process was repeated for each of the participants.

\begin{figure}[h]
\begin{center}
\includegraphics[width=0.8\linewidth]{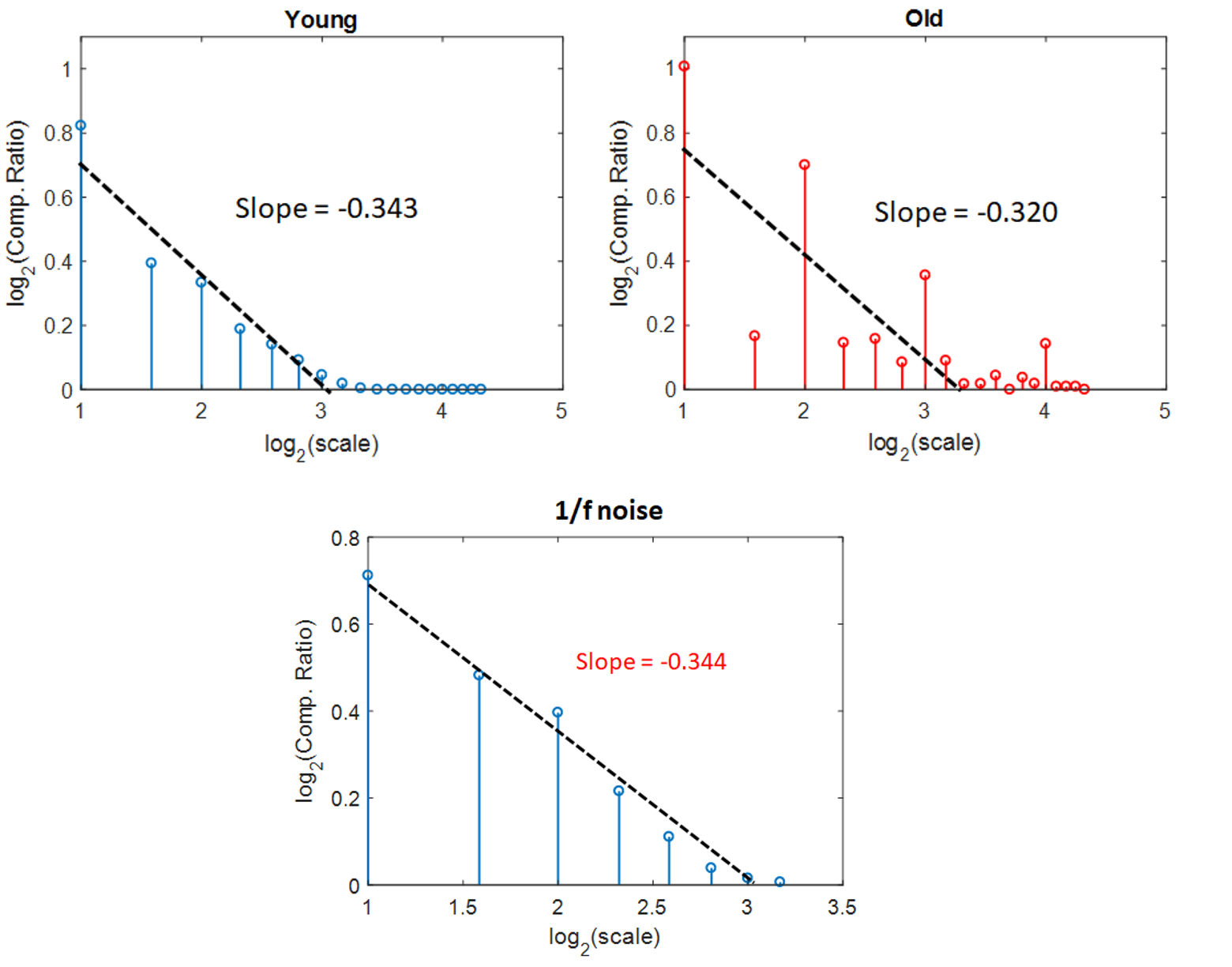}\\
\caption{Compression spectra (in log-log) for real-world data set (`Physionet: Fantasia database'~\cite{goldberger2000physiobank}) and $1/f$ noise. (a) Top left: Spectra of heart interbeat intervals of a `Young' subject~\cite{goldberger2000physiobank}.  (b) Top right:  Spectra of heart interbeat intervals of an `Old' subject~\cite{goldberger2000physiobank}. (c) Bottom: Spectra of simulated $1/f$ noise (of length $10^6$ samples).
  }
\label{fig3}
\end{center}
\end{figure}

Many past studies have shown that the interbeat intervals of a healthy young human heart have fractal-like characteristics~\cite{perkiomaki2005fractal, pikkujamsa1999cardiac, tapanainen2002fractal, iyengar1996age, dawi2022complexity}. We wanted to analyse whether these fractal characteristics are also visible on the compression spectrum. The compression spectra of interbeat RR intervals on log-log scale for a young and an elderly adult are plotted in Fig.~\ref{fig3}(a) and Fig.~\ref{fig3}(b) respectively. 

It has been shown in~\cite{iyengar1996age} that the power spectral analysis of RR time series for young subjects displays long range correlation akin to $1/f$ noise. On the other hand, for elderly subjects, the fractal scaling properties are disturbed as shown in the same study. This may be a result of degradation of heart's physiology with age. Using compression spectrum, we observe something similar. Young adult's spectra as seen in Fig.~\ref{fig3}(a) is fractal like as a good linear fit can be made to the points in the spectrum. Additionally, the slope of this fit ($=-0.343$) is very close to slope of compression spectra seen in case of $1/f$ noise ($=-0.344$). For the elderly subject, we get a poor linear fit and a slope which is not similar to that of $1/f$ noise. Further extensive analysis and study is warranted before we can provide conclusive results. 

%

\section{Discussion and Conclusions}
Spectral analysis based on compressibility can be a useful tool for nonlinear time series analysis where traditional Fourier analysis has limitations. It can provide insight into the information content of signals at different scales without the need to filter the signals. We have, for the first time, proposed {\it Compression Spectrum} -- a signal analysis and visualization tool that attempts to bridge both worlds -- Signal Processing and Information Theory.

Simulation study shows that the compression spectrum gives meaningful and interpretable representation of some basic signals (such as periodic, sinusoidal, chaotic and random) considered. A notion of {\it bandwidth} - defined as the number of scales which exhibit a non-zero compression ratio, is also proposed.  However, the reason why weak chaos shows higher bandwidth compared to strong chaos needs further investigation. One reason could be that under weak chaos, a few attracting periodic orbits skew the distribution of patterns at various scales thereby making them less random (implying certain patterns at certain scales are differentially preferred). As randomness increases~\footnote{This is also the reason why strong chaos and robust chaos have lower bandwidth than weak chaos. The lack of attracting orbits at $a=4.0$ for the logistic map makes the trajectory essentially random-like with no prefered pattern at any scale.}, the information content is concentrated in smaller scales (patterns that arise out of pure chance) with hardly anything in the higher scales (no preferred pattern at any scale whatsoever). Thus, lack of randomness and periodicity  is one way to achieve higher bandwidth.  This is likely to be best achieved by weak chaos. 

The relationship between the slope of compression spectrum (linear fit to bandwidth) to that of the frequency spectrum demands further exploration. Multiple lines fit to the compression spectrum may indicate multifractal nature of processes. This aspect will be studied using some simulations in future work. If destruction of fractal characteristics of compression spectrum of RR interval of elderly subjects can provide any additional information compared to regular frequency spectrum will also be studied in the future.

Preliminary analysis shows that the proposed compression spectrum seems to exhibit analogous properties to those of Fourier transform such as {\it leakage} and {\it harmonics}. These will also be discussed exhaustively in future work.

\section*{Code Availability}
A \textsc{Matlab} program of the Compression Spectrum (along with a few examples) for free download and public use (for research purposes only) can be found here: \url{https://sites.google.com/site/nithinnagaraj2/software_download}.

\begin{acknowledgments}
A.K. is grateful to the financial support provided by Czech Science Foundation, Project No.~GA19-16066S and by the Czech Academy of Sciences, Praemium Academiae awarded to M. Palu\v{s}.
\end{acknowledgments}


\bibliography{apssamp_refs}

\begin{thebibliography}{15}%
\makeatletter
\providecommand \@ifxundefined [1]{%
 \@ifx{#1\undefined}
}%
\providecommand \@ifnum [1]{%
 \ifnum #1\expandafter \@firstoftwo
 \else \expandafter \@secondoftwo
 \fi
}%
\providecommand \@ifx [1]{%
 \ifx #1\expandafter \@firstoftwo
 \else \expandafter \@secondoftwo
 \fi
}%
\providecommand \natexlab [1]{#1}%
\providecommand \enquote  [1]{``#1''}%
\providecommand \bibnamefont  [1]{#1}%
\providecommand \bibfnamefont [1]{#1}%
\providecommand \citenamefont [1]{#1}%
\providecommand \href@noop [0]{\@secondoftwo}%
\providecommand \href [0]{\begingroup \@sanitize@url \@href}%
\providecommand \@href[1]{\@@startlink{#1}\@@href}%
\providecommand \@@href[1]{\endgroup#1\@@endlink}%
\providecommand \@sanitize@url [0]{\catcode `\\12\catcode `\$12\catcode
  `\&12\catcode `\#12\catcode `\^12\catcode `\_12\catcode `\%12\relax}%
\providecommand \@@startlink[1]{}%
\providecommand \@@endlink[0]{}%
\providecommand \url  [0]{\begingroup\@sanitize@url \@url }%
\providecommand \@url [1]{\endgroup\@href {#1}{\urlprefix }}%
\providecommand \urlprefix  [0]{URL }%
\providecommand \Eprint [0]{\href }%
\providecommand \doibase [0]{https://doi.org/}%
\providecommand \selectlanguage [0]{\@gobble}%
\providecommand \bibinfo  [0]{\@secondoftwo}%
\providecommand \bibfield  [0]{\@secondoftwo}%
\providecommand \translation [1]{[#1]}%
\providecommand \BibitemOpen [0]{}%
\providecommand \bibitemStop [0]{}%
\providecommand \bibitemNoStop [0]{.\EOS\space}%
\providecommand \EOS [0]{\spacefactor3000\relax}%
\providecommand \BibitemShut  [1]{\csname bibitem#1\endcsname}%
\let\auto@bib@innerbib\@empty
\bibitem [{\citenamefont {Oppenheim}\ and\ \citenamefont
  {Schafer}(1975)}]{oppenheim1975digital}%
  \BibitemOpen
  \bibfield  {author} {\bibinfo {author} {\bibfnamefont {A.~V.}\ \bibnamefont
  {Oppenheim}}\ and\ \bibinfo {author} {\bibfnamefont {R.~W.}\ \bibnamefont
  {Schafer}},\ }\bibfield  {title} {\bibinfo {title} {Digital signal
  processing(book)},\ }\href@noop {} {\bibfield  {journal} {\bibinfo  {journal}
  {Research supported by the Massachusetts Institute of Technology, Bell
  Telephone Laboratories, and Guggenheim Foundation. Englewood Cliffs, N. J.,
  Prentice-Hall, Inc., 1975. 598 p}\ } (\bibinfo {year} {1975})}\BibitemShut
  {NoStop}%
\bibitem [{\citenamefont {Shannon}(1948)}]{shannon1948mathematical}%
  \BibitemOpen
  \bibfield  {author} {\bibinfo {author} {\bibfnamefont {C.~E.}\ \bibnamefont
  {Shannon}},\ }\bibfield  {title} {\bibinfo {title} {A mathematical theory of
  communication},\ }\href@noop {} {\bibfield  {journal} {\bibinfo  {journal}
  {The Bell system technical journal}\ }\textbf {\bibinfo {volume} {27}},\
  \bibinfo {pages} {379} (\bibinfo {year} {1948})}\BibitemShut {NoStop}%
\bibitem [{\citenamefont {Cover}(1999)}]{cover1999elements}%
  \BibitemOpen
  \bibfield  {author} {\bibinfo {author} {\bibfnamefont {T.~M.}\ \bibnamefont
  {Cover}},\ }\href@noop {} {\emph {\bibinfo {title} {Elements of information
  theory}}}\ (\bibinfo  {publisher} {John Wiley \& Sons},\ \bibinfo {year}
  {1999})\BibitemShut {NoStop}%
\bibitem [{\citenamefont {Sayood}(2017)}]{sayood2017introduction}%
  \BibitemOpen
  \bibfield  {author} {\bibinfo {author} {\bibfnamefont {K.}~\bibnamefont
  {Sayood}},\ }\href@noop {} {\emph {\bibinfo {title} {Introduction to data
  compression}}}\ (\bibinfo  {publisher} {Morgan Kaufmann},\ \bibinfo {year}
  {2017})\BibitemShut {NoStop}%
\bibitem [{\citenamefont {Nagaraj}\ \emph {et~al.}(2013)\citenamefont
  {Nagaraj}, \citenamefont {Balasubramanian},\ and\ \citenamefont
  {Dey}}]{nagaraj2013new}%
  \BibitemOpen
  \bibfield  {author} {\bibinfo {author} {\bibfnamefont {N.}~\bibnamefont
  {Nagaraj}}, \bibinfo {author} {\bibfnamefont {K.}~\bibnamefont
  {Balasubramanian}},\ and\ \bibinfo {author} {\bibfnamefont {S.}~\bibnamefont
  {Dey}},\ }\bibfield  {title} {\bibinfo {title} {A new complexity measure for
  time series analysis and classification},\ }\href@noop {} {\bibfield
  {journal} {\bibinfo  {journal} {The European Physical Journal Special
  Topics}\ }\textbf {\bibinfo {volume} {222}},\ \bibinfo {pages} {847}
  (\bibinfo {year} {2013})}\BibitemShut {NoStop}%
\bibitem [{\citenamefont {Nagaraj}\ and\ \citenamefont
  {Balasubramanian}(2017)}]{nagaraj2017three}%
  \BibitemOpen
  \bibfield  {author} {\bibinfo {author} {\bibfnamefont {N.}~\bibnamefont
  {Nagaraj}}\ and\ \bibinfo {author} {\bibfnamefont {K.}~\bibnamefont
  {Balasubramanian}},\ }\bibfield  {title} {\bibinfo {title} {Three
  perspectives on complexity: entropy, compression, subsymmetry},\ }\href@noop
  {} {\bibfield  {journal} {\bibinfo  {journal} {The European Physical Journal
  Special Topics}\ }\textbf {\bibinfo {volume} {226}},\ \bibinfo {pages} {3251}
  (\bibinfo {year} {2017})}\BibitemShut {NoStop}%
\bibitem [{\citenamefont {Ebeling}\ and\ \citenamefont
  {Jim{\'e}nez-Monta{\~n}o}(1980)}]{ebeling1980grammars}%
  \BibitemOpen
  \bibfield  {author} {\bibinfo {author} {\bibfnamefont {W.}~\bibnamefont
  {Ebeling}}\ and\ \bibinfo {author} {\bibfnamefont {M.~A.}\ \bibnamefont
  {Jim{\'e}nez-Monta{\~n}o}},\ }\bibfield  {title} {\bibinfo {title} {On
  grammars, complexity, and information measures of biological
  macromolecules},\ }\href@noop {} {\bibfield  {journal} {\bibinfo  {journal}
  {Mathematical Biosciences}\ }\textbf {\bibinfo {volume} {52}},\ \bibinfo
  {pages} {53} (\bibinfo {year} {1980})}\BibitemShut {NoStop}%
\bibitem [{Note1()}]{Note1}%
  \BibitemOpen
  \bibinfo {note} {A symbolic sequence is a sequence consisting of symbols for
  each value of the time series, obtained by quantization. For example, a
  binary symbolic sequence would be one where every value of the time-series is
  replaced by `$0$' or `$1$' based on some threshold. One can have any finite
  number of bins for quantizing the input time series. Bins could be uniform or
  not. Typically, uniform bins are chosen to create the symbolic sequence from
  the time-series.}\BibitemShut {Stop}%
\bibitem [{\citenamefont {Goldberger}\ \emph {et~al.}(2000)\citenamefont
  {Goldberger}, \citenamefont {Amaral}, \citenamefont {Glass}, \citenamefont
  {Hausdorff}, \citenamefont {Ivanov}, \citenamefont {Mark}, \citenamefont
  {Mietus}, \citenamefont {Moody}, \citenamefont {Peng},\ and\ \citenamefont
  {Stanley}}]{goldberger2000physiobank}%
  \BibitemOpen
  \bibfield  {author} {\bibinfo {author} {\bibfnamefont {A.~L.}\ \bibnamefont
  {Goldberger}}, \bibinfo {author} {\bibfnamefont {L.~A.}\ \bibnamefont
  {Amaral}}, \bibinfo {author} {\bibfnamefont {L.}~\bibnamefont {Glass}},
  \bibinfo {author} {\bibfnamefont {J.~M.}\ \bibnamefont {Hausdorff}}, \bibinfo
  {author} {\bibfnamefont {P.~C.}\ \bibnamefont {Ivanov}}, \bibinfo {author}
  {\bibfnamefont {R.~G.}\ \bibnamefont {Mark}}, \bibinfo {author}
  {\bibfnamefont {J.~E.}\ \bibnamefont {Mietus}}, \bibinfo {author}
  {\bibfnamefont {G.~B.}\ \bibnamefont {Moody}}, \bibinfo {author}
  {\bibfnamefont {C.-K.}\ \bibnamefont {Peng}},\ and\ \bibinfo {author}
  {\bibfnamefont {H.~E.}\ \bibnamefont {Stanley}},\ }\bibfield  {title}
  {\bibinfo {title} {Physiobank, physiotoolkit, and physionet: components of a
  new research resource for complex physiologic signals},\ }\href@noop {}
  {\bibfield  {journal} {\bibinfo  {journal} {circulation}\ }\textbf {\bibinfo
  {volume} {101}},\ \bibinfo {pages} {e215} (\bibinfo {year}
  {2000})}\BibitemShut {NoStop}%
\bibitem [{\citenamefont {Iyengar}\ \emph {et~al.}(1996)\citenamefont
  {Iyengar}, \citenamefont {Peng}, \citenamefont {Morin}, \citenamefont
  {Goldberger},\ and\ \citenamefont {Lipsitz}}]{iyengar1996age}%
  \BibitemOpen
  \bibfield  {author} {\bibinfo {author} {\bibfnamefont {N.}~\bibnamefont
  {Iyengar}}, \bibinfo {author} {\bibfnamefont {C.}~\bibnamefont {Peng}},
  \bibinfo {author} {\bibfnamefont {R.}~\bibnamefont {Morin}}, \bibinfo
  {author} {\bibfnamefont {A.~L.}\ \bibnamefont {Goldberger}},\ and\ \bibinfo
  {author} {\bibfnamefont {L.~A.}\ \bibnamefont {Lipsitz}},\ }\bibfield
  {title} {\bibinfo {title} {Age-related alterations in the fractal scaling of
  cardiac interbeat interval dynamics},\ }\href@noop {} {\bibfield  {journal}
  {\bibinfo  {journal} {American Journal of Physiology-Regulatory, Integrative
  and Comparative Physiology}\ }\textbf {\bibinfo {volume} {271}},\ \bibinfo
  {pages} {R1078} (\bibinfo {year} {1996})}\BibitemShut {NoStop}%
\bibitem [{\citenamefont {Perki{\"o}m{\"a}ki}\ \emph
  {et~al.}(2005)\citenamefont {Perki{\"o}m{\"a}ki}, \citenamefont
  {M{\"a}kikallio},\ and\ \citenamefont {Huikuri}}]{perkiomaki2005fractal}%
  \BibitemOpen
  \bibfield  {author} {\bibinfo {author} {\bibfnamefont {J.~S.}\ \bibnamefont
  {Perki{\"o}m{\"a}ki}}, \bibinfo {author} {\bibfnamefont {T.~H.}\ \bibnamefont
  {M{\"a}kikallio}},\ and\ \bibinfo {author} {\bibfnamefont {H.~V.}\
  \bibnamefont {Huikuri}},\ }\bibfield  {title} {\bibinfo {title} {Fractal and
  complexity measures of heart rate variability},\ }\href@noop {} {\bibfield
  {journal} {\bibinfo  {journal} {Clinical and experimental hypertension}\
  }\textbf {\bibinfo {volume} {27}},\ \bibinfo {pages} {149} (\bibinfo {year}
  {2005})}\BibitemShut {NoStop}%
\bibitem [{\citenamefont {Pikkuj{\"a}ms{\"a}}\ \emph
  {et~al.}(1999)\citenamefont {Pikkuj{\"a}ms{\"a}}, \citenamefont
  {M{\"a}kikallio}, \citenamefont {Sourander}, \citenamefont {R{\"a}ih{\"a}},
  \citenamefont {Puukka}, \citenamefont {Skytt{\"a}}, \citenamefont {Peng},
  \citenamefont {Goldberger},\ and\ \citenamefont
  {Huikuri}}]{pikkujamsa1999cardiac}%
  \BibitemOpen
  \bibfield  {author} {\bibinfo {author} {\bibfnamefont {S.~M.}\ \bibnamefont
  {Pikkuj{\"a}ms{\"a}}}, \bibinfo {author} {\bibfnamefont {T.~H.}\ \bibnamefont
  {M{\"a}kikallio}}, \bibinfo {author} {\bibfnamefont {L.~B.}\ \bibnamefont
  {Sourander}}, \bibinfo {author} {\bibfnamefont {I.~J.}\ \bibnamefont
  {R{\"a}ih{\"a}}}, \bibinfo {author} {\bibfnamefont {P.}~\bibnamefont
  {Puukka}}, \bibinfo {author} {\bibfnamefont {J.}~\bibnamefont {Skytt{\"a}}},
  \bibinfo {author} {\bibfnamefont {C.-K.}\ \bibnamefont {Peng}}, \bibinfo
  {author} {\bibfnamefont {A.~L.}\ \bibnamefont {Goldberger}},\ and\ \bibinfo
  {author} {\bibfnamefont {H.~V.}\ \bibnamefont {Huikuri}},\ }\bibfield
  {title} {\bibinfo {title} {Cardiac interbeat interval dynamics from childhood
  to senescence: comparison of conventional and new measures based on fractals
  and chaos theory},\ }\href@noop {} {\bibfield  {journal} {\bibinfo  {journal}
  {Circulation}\ }\textbf {\bibinfo {volume} {100}},\ \bibinfo {pages} {393}
  (\bibinfo {year} {1999})}\BibitemShut {NoStop}%
\bibitem [{\citenamefont {Tapanainen}\ \emph {et~al.}(2002)\citenamefont
  {Tapanainen}, \citenamefont {Thomsen}, \citenamefont {K{\o}ber},
  \citenamefont {Torp-Pedersen}, \citenamefont {M{\"a}kikallio}, \citenamefont
  {Still}, \citenamefont {Lindgren},\ and\ \citenamefont
  {Huikuri}}]{tapanainen2002fractal}%
  \BibitemOpen
  \bibfield  {author} {\bibinfo {author} {\bibfnamefont {J.~M.}\ \bibnamefont
  {Tapanainen}}, \bibinfo {author} {\bibfnamefont {P.~E.~B.}\ \bibnamefont
  {Thomsen}}, \bibinfo {author} {\bibfnamefont {L.}~\bibnamefont {K{\o}ber}},
  \bibinfo {author} {\bibfnamefont {C.}~\bibnamefont {Torp-Pedersen}}, \bibinfo
  {author} {\bibfnamefont {T.~H.}\ \bibnamefont {M{\"a}kikallio}}, \bibinfo
  {author} {\bibfnamefont {A.-M.}\ \bibnamefont {Still}}, \bibinfo {author}
  {\bibfnamefont {K.~S.}\ \bibnamefont {Lindgren}},\ and\ \bibinfo {author}
  {\bibfnamefont {H.~V.}\ \bibnamefont {Huikuri}},\ }\bibfield  {title}
  {\bibinfo {title} {Fractal analysis of heart rate variability and mortality
  after an acute myocardial infarction},\ }\href@noop {} {\bibfield  {journal}
  {\bibinfo  {journal} {The American journal of cardiology}\ }\textbf {\bibinfo
  {volume} {90}},\ \bibinfo {pages} {347} (\bibinfo {year} {2002})}\BibitemShut
  {NoStop}%
\bibitem [{\citenamefont {Dawi}\ \emph {et~al.}(2022)\citenamefont {Dawi},
  \citenamefont {Maresova},\ and\ \citenamefont {Namazi}}]{dawi2022complexity}%
  \BibitemOpen
  \bibfield  {author} {\bibinfo {author} {\bibfnamefont {N.~M.}\ \bibnamefont
  {Dawi}}, \bibinfo {author} {\bibfnamefont {P.}~\bibnamefont {Maresova}},\
  and\ \bibinfo {author} {\bibfnamefont {H.}~\bibnamefont {Namazi}},\
  }\bibfield  {title} {\bibinfo {title} {Complexity-based analysis of heart
  rate variability during aging},\ }\href@noop {} {\bibfield  {journal}
  {\bibinfo  {journal} {Fractals}\ }\textbf {\bibinfo {volume} {30}},\ \bibinfo
  {pages} {2250198} (\bibinfo {year} {2022})}\BibitemShut {NoStop}%
\bibitem [{Note2()}]{Note2}%
  \BibitemOpen
  \bibinfo {note} {This is also the reason why strong chaos and robust chaos
  have lower bandwidth than weak chaos. The lack of attracting orbits at
  $a=4.0$ for the logistic map makes the trajectory essentially random-like
  with no prefered pattern at any scale.}\BibitemShut {Stop}%
\end{thebibliography}%

\end{document}